\documentclass[conference,letterpaper]{IEEEtran}
\IEEEoverridecommandlockouts
\usepackage{cite}
\usepackage{amsmath,amssymb,amsfonts}
\usepackage{algorithmic}
\usepackage[export]{adjustbox}
\usepackage{graphicx}
\usepackage{subcaption}
\usepackage{textcomp}
\usepackage{tabularx}
\usepackage[hidelinks]{hyperref}
\usepackage[layout={index,margin}]{fixme}
\usepackage{booktabs}
\usepackage{multirow}
\usepackage[table,xcdraw]{xcolor}
\def\BibTeX{{\rm B\kern-.05em{\sc i\kern-.025em b}\kern-.08em
    T\kern-.1667em\lower.7ex\hbox{E}\kern-.125emX}}

\begin{document}
\title{Detecting Cloud-Based Phishing Attacks by Combining Deep Learning Models}

\author{\IEEEauthorblockN{Birendra Jha}
\IEEEauthorblockA{\textit{Eydle Inc.}}
\and
\IEEEauthorblockN{Medha Atre}
\IEEEauthorblockA{\textit{Eydle Inc.}}
\and
\IEEEauthorblockN{Ashwini Rao}
\IEEEauthorblockA{\textit{Eydle Inc.}}
}



\maketitle

\begin{abstract}
Web-based phishing attacks nowadays exploit popular cloud web hosting services and apps such as Google Sites and Typeform for hosting their attacks. Since these attacks originate from reputable domains and IP addresses of the cloud services, traditional phishing detection methods such as IP reputation monitoring and blacklisting are not very effective. Here we investigate the effectiveness of deep learning models in detecting this class of cloud-based phishing attacks. Specifically, we evaluate deep learning models for three phishing detection methods--LSTM model for URL analysis, YOLOv2 model for logo analysis, and triplet network model for visual similarity analysis. We train the models using well-known datasets and test their performance on cloud-based phishing attacks in the wild. Our results qualitatively explain why the models succeed or fail. Furthermore, our results highlight how combining results from the individual models can improve the effectiveness of detecting cloud-based phishing attacks.
\end{abstract}

\begin{IEEEkeywords}
Deep Learning, Phishing, Cybersecurity
\end{IEEEkeywords}

\section{Introduction} \label{sec:intro}

Phishing continues to be a top cyber crime~\cite{fbi:report, proofpoint}. The US Federal Bureau of Investigation reported that phishing attacks nearly doubled in 2020~\cite{fbi:report}.  
In 2020, there was also a rise in the use of cloud-based services such as Google Drive and Office 365 to host phishing sites~\cite{proofpoint}. Popular cloud-based web hosting services and apps have reputed domains and IP addresses. Hence, phishing detection methods using IP reputation monitoring are not very effective against attacks hosted on these services. Blacklisting  IP addresses and domains of such cloud services is difficult. For example, in 2018, Instagram blocked Linktree, a social media reference landing page service, citing spam and phishing originating from Linktree~\cite{LinktreeInstagram}. Later Instagram had to revoke the ban. 

In this work, we explore the effectiveness of deep learning models in detecting phishing attacks hosted on well-known cloud-based services. We investigate deep learning models for three phishing detection methods: URL analysis, visual similarity analysis and logo detection. First, we evaluate Long-Short-Term-Memory (LSTM) model for detecting whether a given URL is a phishing URL. Second, we apply a Triplet network visual similarity model to detect whether a phishing webpage is visually similar to a brand's webpage. Lastly, we use You-Only-Look-Once (YOLO) object detection model for detecting brand logos present in a webpage. Logo detection helps in identifying the brand being phished.

In addition to investigating the effectiveness of individual deep learning models, we explore whether combining the outputs of individual models can improve the effectiveness of detecting cloud-based phishing attacks. For instance, could combining outputs help us flag false negatives that would miss phishing URLs or webpages? To evaluate our hypotheses about individual models and combining outputs, we train the models on well-known data sets including Kaggle URL dataset~\cite{kaggle:phishing}, Flickr47 logo dataset~\cite{flickr47} and VisualPhishNet dataset~\cite{visualphishnet}. We test model performance on phishing attacks in the wild. To do so, we use a dataset collected from PhishTank~\cite{phishtank}, a community site for sharing latest phishing data. Our results qualitatively explain why the models succeed or fail individually. They also explain how combining outputs can improve the effectiveness of detecting cloud-based phishing attacks. To guide the reader through our analyses, we use five running examples listed in Table~\ref{tab:runningexamples}. These examples taken from our PhishTank test dataset illustrate real-life cloud-based phishing attacks. Our main contributions are as follows. 
\begin{itemize}
\item We evaluate effectiveness of three existing deep learning models in detecting cloud-based phishing attacks. Our qualitative analysis shows why models succeed or fail.
\item We evaluate whether combining the outputs of the three individual deep learning models can help us detect cloud-based phishing attacks more accurately. Our results provide early evidence that support our hypothesis. 
\end{itemize}

\begin{table}[t]
\resizebox{\columnwidth}{!}{%
\begin{tabular}{@{}llll@{}}
\toprule
ID & Brand Targeted & Phishing URL                                        & Service     \\ \midrule
G1 & Google         & https://sites.google.com/view/yaho000/home    & Google Sites      \\
G2 & Google         & https://sites.google.com/view/fgjdfghduhdxuxu/home & Google Sites \\
B1 & British Telecom             & https://dfghhgdsdf.weebly.com             & Weebly       \\
B2 & British Telecom             & https://ofifice.weebly.com                 & Weebly       \\
D1 & DHL            & https://dhmpxmsb6lk.typeform.com/to/h99lvret                      & Typeform     \\ \bottomrule
\end{tabular}%
}
\caption{Illustrative Running Examples of Cloud-based Phishing Attacks}
\label{tab:runningexamples}
\end{table}

\section{Background and Related Work} \label{sec:relwork}
Compared to the existing work, we focus on cloud-based phishing attacks hosted on popular cloud services and apps. We discuss background and work related to analysis methods and deep learning models that we use for phishing detection.
\subsubsection*{URL Analysis} \label{sec:urldetectrelwork}
Broadly, methods for detecting phishing URLs use lexical (URL length, domain name, special characters, etc.) or host-based (IP, host location, registration date, etc.) features of a URL. Machine learning models for detecting phishing URLs use hand-picked features~\cite{miyamoto2008evaluation} either lexical alone~\cite{blum2010lexical, zouina2017novel} or a combination of lexical and other features such as site rank~\cite{nguyen2014commantel}. Researchers have also used deep learning models to detect phishing URLs~\cite{woodbridge2018detecting, ozcan2021hybrid}. One advantage of deep learning models is that they learn the best features automatically from the data, and can adapt as attacks evolve. Compared to the existing work, we specifically investigate URLs of cloud-based phishing attacks. For this type of attack, we expect models that use lexical features e.g., domain name~\cite{zouina2017novel,ozcan2021hybrid} to be impacted. Similarly, we expect impact on the performance of models that use host information e.g., IP address. We investigate a lightweight Long-Short-Term Memory (LSTM) model based on lexical features~\cite{csirgadget:blog, slalom:medium}.   

\subsubsection*{Visual Similarity Analysis} \label{sec:visualrelwork}
In contrast to URL analysis, content analysis methods detect phishing by  
analyzing the content of a webpage. They can extract textual features such as title and keywords~\cite{zhang2007cantina54, DomainNameBased_URLlength2018}, code such as HTML, DOM or CSS~\cite{rosiello2007layout41, mao2017phishing33}, or images. Visual similarity methods analyze a webpage rendered as an image~\cite{lam2009counteracting25, fu2006detecting14, visualphishnet}. Hybrid methods can combine text and image analysis~\cite{zhang2011textual53, van2021combining}. These methods compute a similarity score between a suspected phishing page and a legitimate webpage. For cloud-based phishing attacks, we expect the impact on visual similarity methods to be smaller than that on URL methods because the attackers try to make the page appear similar to the target being phished even for cloud-based phishing attacks. We use a triplet network deep learning model called VisualPhishNet~\cite{visualphishnet} to do visual similarity analysis.

\subsubsection*{Logo Detection} \label{sec:logodetectrelwork}
A logo detection method cannot independently identify phishing, but it can be used alongside other methods to detect the brand being phished~\cite{afroz2011phishzoo1, bozkir2020logosense,lin2021phishpedia}. For this purpose, researchers have used traditional computer vision methods such as Scale-Invariant Feature Transform (SIFT)~\cite{afroz2011phishzoo1} and Histogram of Oriented Gradients (HOG)~\cite{bozkir2020logosense}, and deep learning methods such as Faster Regional-CNN~\cite{lin2021phishpedia}.
Extracting a logo from webpage segmentation and finding the matching logo using the content-based image retrieval mechanism of the Google Image Search engine is another method~\cite{chang2013phishing6}.
Traditional methods can be slower and less accurate~\cite{severo2018benchmark} than deep learning methods such as YOLO~\cite{eldho2019yolo,yang2019fast}, Single-Shot Detection (SSD)~\cite{su2020scalable} and Faster R-CNN~\cite{ren2015fasterRCNN}. We use a YOLO deep learning model~\cite{yolo9000,akarsh:yolo2}. 



\subsubsection*{Ensemble Detection} \label{sec:relworkensemble}
To improve phishing detection accuracy, researchers have used ensemble methods to combine multiple machine learning models~\cite{zhuang2012intelligent, ubing2019phishing, basit2020novel, al2021optimized, subasi2020comparison} and multiple deep learning models~\cite{vecliuc2021experimental, ozcan2021hybrid,lin2021phishpedia}. 
These methods can combine heterogeneous or homogeneous models. Simple methods combine outputs of different models using statistical summary techniques such as mean and weighted average. Advanced methods can use the outputs of base models to train a higher meta model. Sequential methods train models in sequence and parallel methods train models in parallel. We use three heterogeneous deep learning models in parallel. We do not combine the outputs in a traditional sense, but examine them jointly to understand how multiple models can improve phishing detection accuracy for cloud-based phishing attacks.

Most of the ensemble methods rely on specific features extracted from the URLs or websites. In contrary, we use deep neural networks to extract the relevant features automatically during model training. In~\cite{vecliuc2021experimental}, authors use CNN-based text classification model for URL, Support Vector Machine (SVM) model for processing TF-IDF representation of the HTML of the webpage, and YOLOv3-tiny model for logo and copyright detection in the screenshot of the page. A k-Nearest-Neighbour (kNN) ensemble model is used to combine the results of the three models and classify the URL as phishing or legitimate. For training the models, authors  used their in-house manually labeled datasets, which are not available publicly. In general, there are no publicly available datasets of phishing URLs and the corresponding pages in HTML and image (screenshot) format. We trained our URL and logo models on publicly available, widely used datasets and, for our visual similarity model, we used VisualPhishNet's~\cite{visualphishnet} training dataset. 

\section{Design} \label{sec:design}
We specifically use deep learning models because of their ability to automatically extract and learn the best features for classification and object detection tasks. To detect phishing, one can use different types of information related to a website such as IP address, URL, source code, logos, visual appearance and text. For cloud-based phishing attacks, IP reputation monitoring is not very effective because popular cloud services have good IP reputation. Here, we focus on deep learning models that use URL, visual similarity and logo information to detect phishing. In the future, we plan to investigate the use of additional information. 

We selected one of our models to be URL analysis because it can be faster and more efficient compared to other approaches. Domain name is part of the URL, and, for cloud-based phishing attacks, we expect it to impact phishing detection accuracy of the URL model. Visual similarity analysis is useful because phishing websites look similar to the respective legitimate websites, and deep learning models are particularly well suited for visual analysis tasks. Because we analyze the phishing webpage as an image, we do not expect the detection accuracy of the model to change for cloud-based phishing attacks. We choose logo detection to identify the brand targeted by the attacker. By itself, logo detection does not identify phishing. Logo detection is a visual analysis task and appropriate for deep learning models.

In addition to investigating how individual deep learning models perform at cloud-based phishing detection, we examine their outputs jointly and identify how combining the outputs of individual models can improve detection accuracy. Below we discuss our choice of deep learning model for the three phishing analysis methods. We discuss the datasets we use to train and test the models.
Although we focus here on cloud-based phishing attacks, our models and methods can work on non-cloud based phishing URLs too, which are included in the training and testing of the models.


\subsection{LSTM Model for URL Analysis} \label{sec:urlmodel}

LSTM models are well-suited for learning tasks where \textit{sequence} matters.
Because a URL is a sequence of alphabets, numeric digits and other symbols, LSTM models can be used to detect phishing URLs~\cite{ozcan2021hybrid}. 
An advantage of using a LSTM model is that the learning capacity can be tweaked by changing the size of the network, i.e., the number of LSTM layers, number of nodes per layer, and the number of epochs used for training the model, without the need of manually tweaking or changing the features every time an attack evolves. This may not be true for machine learning models trained on hand-picked features of a given URL training dataset.
We explore the performance of a LSTM model that is lightweight in terms of computational cost~\cite{csirgadget:blog, slalom:medium}. More complex LSTM models could perform better, but at higher computational cost. Our LSTM model consists of embedding, dropout, LSTM, and dense layers. The embedding layer's output dimension is 32. This layer is followed by a 0.5 dropout layer, an LSTM layer, a 0.5 dropout layer, and a dense layer for final probability score of phishing. To improve the accuracy of the original model, we increased the number of nodes in the LSTM layer from 16 to 64.


We train and validate the LSTM models on a dataset with 482916 URLs of which 380088 are legitimate and 102828 are phishing. The dataset contains 420464 URLs from the Kaggle phishing URL dataset~\cite{kaggle:phishing} and 62452 URLs from OpenPhish phishing database~\cite{csirgadgets}. To compare performance on unseen data, in September 2021, we collected 237 URLs in the wild (117 legitimate \& 120 phishing) from PhishTank.
During March-June 2022, we also collected 455 cloud-based phishing URLs hosted on reputed domains e.g., Google Sites, Weebly, Linktree, Duckdns, Yolasite, Github, and Azure Websites. We use this dataset to test our URL model as well as the other two models. 



\subsection{Triplet Network Model for Visual Similarity Analysis} \label{sec:tripletnet}
To detect cloud-based phishing using visual similarity analysis, we choose a Triplet Network model, called VisualPhishNet~\cite{visualphishnet}. This model quantifies the pixel-based visual similarity in terms of a distance metric between a given webpage and legitimate webpages belonging to a brand. Based on a distance threshold, the given webpage can be classified as a phishing page or a legitimate page. Among the different deep learning models available for visual similarity analysis, we choose VisualPhishNet because it allows us to train on multiple webpages belonging to a brand as opposed to a single webpage. Attackers often create phishing webpages that have overlapping similarity with different webpages of a brand. By training on multiple webpages per brand, the model learns similarities across multiple webpages of a brand and can detect phishing pages with higher accuracy~\cite{visualphishnet}.

A triplet network is an extension of the Siamese network used in facial recognition technology such as FaceNet~\cite{Facenet2015}. The word triplet refers to the triplet loss function~\cite{FacenetLossFunc2005,Facenet2015} used to train the convolutional neural network of the model, which takes three images as inputs---an anchor image of a brand, a positive image of the same brand, and a negative image of a different brand. The loss function is defined in terms of the distance between the anchor and the positive and negative images respectively. The objective is to learn a feature space in which the distance between the anchor and its positive image is smaller than that between the anchor and the negative image. During testing, for a given test image, the model outputs top-k images from the training dataset, which are closest to the test image in terms of the L2 distance between the output image and the test image. 
An image with L2 distance less than a threshold is classified as phishing.


We use the triplet network model architecture from the original VisualPhishNet study~\cite{visualphishnet}. The model uses a VGG16 network~\cite{Simonyan2015VeryDC} with the top fully connected layers replaced by a new convolution layer with 512 filters of size 5$\times$5  and ReLU activation. The final layer is a global max pooling layer. The VGG16 layers are initialized with ImageNet~\cite{Simonyan2015VeryDC} weights and the new layer is initialized with He-normal weights. 

To train the model, we use an updated VisualPhishNet dataset~\cite{visualphishnet}. The original dataset has 9363 pages (screenshots) of 155 brands. Since the publication of the dataset, British Telecom (BT), one of the 155 brands, changed its logo. Hence, we updated the data for BT; we added 45 screenshots of legitimate webpages collected from the BT website, and 31 screenshots of  phishing pages collected from PhishTank. We use the VisualPhishNet test dataset with 1400 pages containing 717 phishing pages that target the brands in the training dataset and 683 legitimate pages of brands not in the training dataset. 
As discussed in Section~\ref{sec:urlmodel}, we also created a test dataset of screenshots of cloud-based phishing URLs collected from PhishTank during March-June 2022.

\subsection{YOLO2 Model for Logo Detection} \label{sec:logomodel}

For logo detection, we choose the YOLO object detection model because it uses global features of the page and does fast prediction with a single computation pass through the network~\cite{yolo9000,eldho2019yolo,yang2019fast}. Training a YOLO model from scratch is computationally expensive. Therefore, we use a pretrained YOLO version 2 (YOLO2) model~\cite{akarsh:yolo2} already trained on Flickr-47 dataset consisting of logos from 47 well-known brands~\cite{flickr47}. The model is composed of Darknet-19 feature extractor layers followed by 12 layers for prediction~\cite{yolo9000}. To detect logos of new brands, we extend the pretrained model using the \textit{transfer learning} technique~\cite{DeepFaceRecog_2015}, which avoids training from scratch. For that we deconstruct the last prediction layer of the model to adjust the dimension of the output.

To do transfer learning on two new brands, Chase and BT, we extend the Flickr47 dataset by adding images containing the brand logos. We collect the images manually from the web and annotate the logos within images using LabelImg~\cite{labelimg}. We extend the training dataset by adding 10 and 107 images for Chase and BT respectively. We extend the test dataset by adding 7 and 11 images for Chase and BT, respectively. The extended Flickr47 dataset contains 49 brands with a total of 1989 training and 1420 test images. To scale to thousands of brands, we can use recently proposed ideas of canonical images~\cite{OSLD_amazon_2019} and automatic generation of training data~\cite{su2020scalable}.

The original YOLO2 model was trained for 10000 epochs. We did transfer learning for 3000 epochs. For the first 2000 epochs, we froze the Darknet-19 backbone by setting the \texttt{trainable} attribute of the respective layers to false. For the following 1000 epochs, we unfroze all the layers including the Darknet-19 layers for fine-tuning. 



\section{Evaluation} \label{sec:experiments}
First we benchmark the performance of URL, visual similarity and logo models. Then, using the running examples in Table~\ref{tab:runningexamples} as a guide, we explain why the individual models succeed or fail at detecting cloud-based phishing attacks. Lastly, we discuss insights from combining model outputs. We trained the URL models on an instance with 8 vCPUs and 65GB RAM, and the visual similarity and logo models on an instance with 32 vCPUs, 128GB RAM and 1 T4 GPU.

\subsection{URL Model} \label{sec:urldetectexpt}


\begin{figure}
\centering
    \includegraphics[width=\columnwidth]{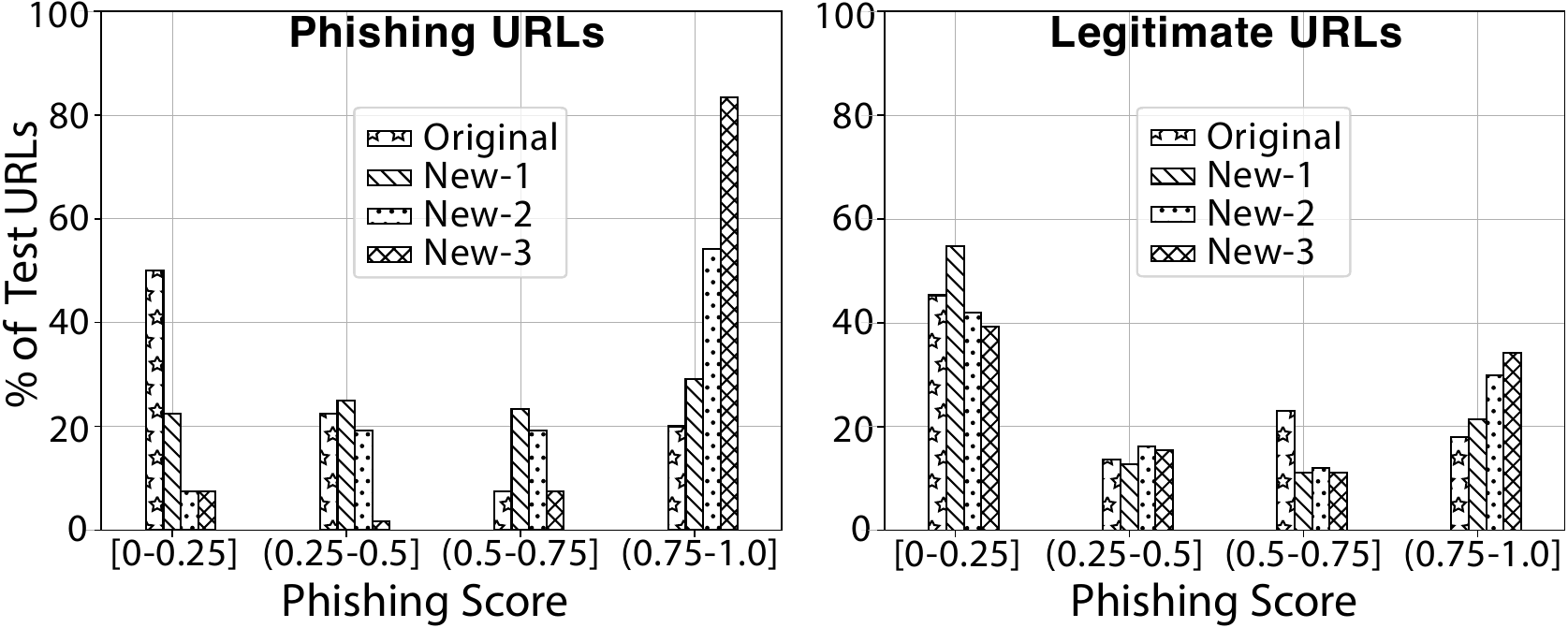}
    \caption{Performance of LSTM URL models on unseen data.}  \label{fig:triophishurl}
\end{figure}
\subsubsection{Model Performance}
We evaluated the performance of the URL model by varying the number of nodes in the LSTM layer and the number of epochs for training the model. We compared four models: Original (16 nodes \& 3 epochs); New-1 (64 nodes \& 3 epochs); New-2 (16 nodes \& 6 epochs); and New-3 (64 nodes \& 6 epochs). We trained each model on our Kaggle + OpenPhish dataset using 0.25 validation split. 
The URL model does not perform a binary classification of a URL as phishing or legitimate. It outputs a phishing score between 0 and 1, where a higher score indicates that the URL is more likely to be a phishing URL. To quantitatively assess the model performance, we bin the phishing scores of a model into four bins, and the number of URL scores falling in a bin is plotted as a bar (Figure~\ref{fig:triophishurl}). 
We can see that New-3 performed the best as it has marked 83\% of the tested phishing URLs with confidence score greater than 0.75.
To compare the performance of the models on unseen data, we used the PhishTank test dataset with URLs in the wild, which consisted of both cloud-based and non-cloud based phishing and legitimate URLs. 
As we can see from Figure~\ref{fig:triophishurl}, New-3 performs best in identifying phishing URLs as phishing, 
because it marks 88\% of the phishing URLs with phishing score greater than 0.75.
However, while identifying legitimate URLs, performance of New-3 and New-1 are comparable. 
Using a conservative approach that minimizes false negatives, we consider New-3 as the best model and trained it for 20 more epochs, which increased its accuracy to 0.96.

\subsubsection{Running Examples}
The URL model identifies two of the five running examples---B1 and D1---as phishing by assigning them high phishing scores. It fails to identify the other three--G1, G2 and B2--as phishing because it assigns them low scores. Our analysis shows that the URL model is more likely to classify a URL as phishing if the URL contains a sequence of random characters that occupies a substantial portion of the URL. For example, B1 contains a random subdomain \textit{dfghhgdsdf}.weebly.com, and the model assigns it a phishing score of 0.997. In contrast, the subdomain in B2 is not random, \textit{ofifice}.weebly.com, and it gets a score of 0.315. G1 and G2 are similar except for one subdirectory--G1 contains \textit{yaho000} and G2 contains \textit{fgjdfghduhdxuxu}. The model assigns G1 a lower score 0.001 and G2 a higher score 0.043. Although G2 contains a random sequence, we believe its score is less compared to B1 because the sequence occupies relatively small portion of the URL. It is likely that the model assigns different weights to different parts of a URL e.g., subdomain vs. subdirectory. Because D1 contains both random subdomain \textit{dhmpxmsb6lk} and subdirectory \textit{h99lvret}, the model assigns it the highest score 0.999. 

The deep learning model automatically learned to use features such as random sequence and URL length to identify phishing. 
The model learned to classify interesting cases where one of the  features may indicate that it is a legitimate URL and another feature may indicate that it a phishing URL, e.g., https://julievoicewireless.weebly.com/ does not have random characters but the first part of the URL, left of the first dot, is relatively long. New-3 gives it a 0.9 score which is correct with it being a phishing URL.
Automatic feature learning is preferable over hand picking features for evolving cloud-based phishing attacks.  A URL model using a hand-picked feature such as the similarity index~\cite{zouina2017novel} assumes that the domain name of the phishing URL is similar to the domain name of the brand it targets. The assumption is flawed in the case of cloud-based phishing attacks. For instance, both B1 and B2 are hosted on weebly.com domain, but they target the bt.com domain. G1 and G2 are hosted on sites.google.com and target google.com, but other phishing attacks hosted on sites.google.com do not target google.com. Hence, the similarity index would lower the model's performance. 

\begin{figure}
\centering
    \includegraphics[width=.7\columnwidth]{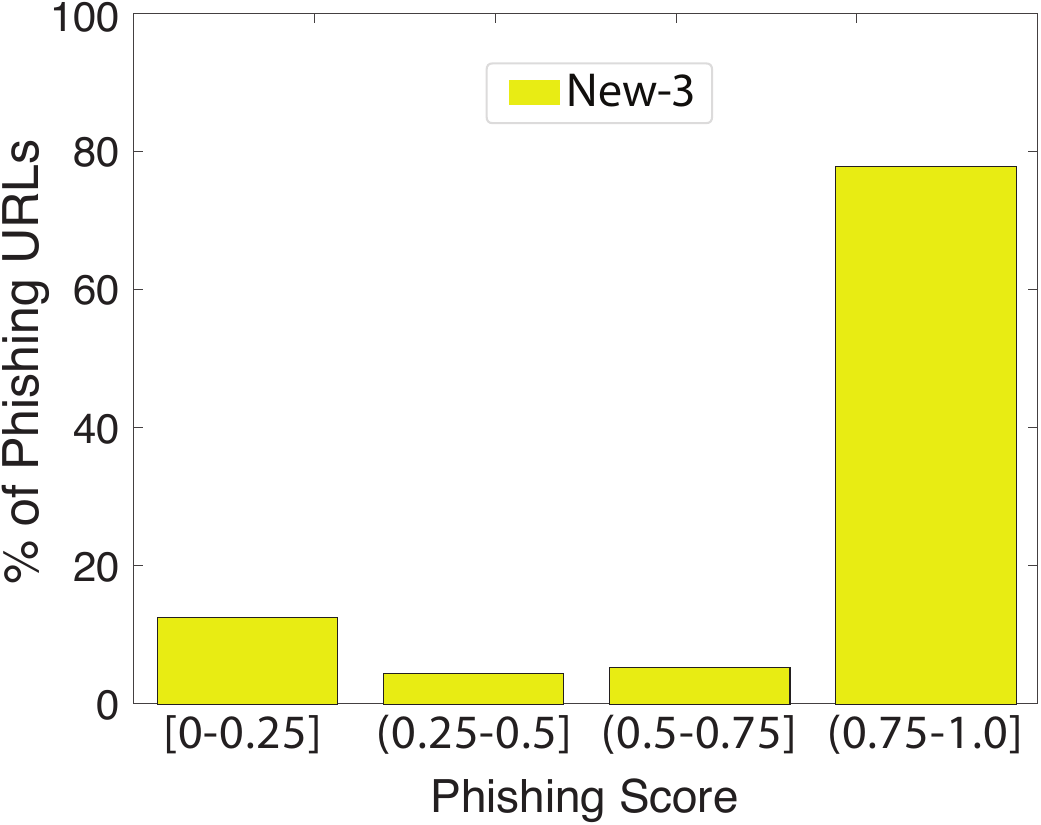}
    \caption{Performance of New-3 URL model on a dataset of 455 cloud-based phishing URLs.}  \label{fig:urlmodel_clouddata}
\end{figure}

The URL model looks for random sequences in a URL. Since the domain name of a popular cloud service is not random and can be a substantial portion of the URL of a cloud-based phishing attack, URLs of cloud-based phishing attacks are less random in appearance than URLs of traditional phishing attacks. Therefore, the model's performance can decrease for cloud-based phishing attacks than for non-cloud based attacks. 
We tested this hypothesis using a dataset of 455 cloud-based phishing URLs from PhishTank that we collected during March-June 2022. The results of the test using the New-3 model are given in Figure~\ref{fig:urlmodel_clouddata}, which shows that 78\% of the cloud-based phishing URLs received scores higher than 0.75  thereby getting classified correctly as phishing. Compared to 88\%  mentioned above for cloud and non-cloud combined, the model performance has decreased on this cloud-based dataset.
This is further supported by the analysis of the false negatives, i.e. URLs with phishing scores less than 0.25 in Figure~\ref{fig:urlmodel_clouddata}. They make up 13\% of the total URLs and they are hosted primarily on Google Sites and Google Docs with a few on Linktree and Weebly domains.
To improve the URL model performance, one can consider using different models to identify the two types of phishing attacks: cloud and non-cloud. If we want to use a single model, it is better to use a hybrid model that combines URL and other information e.g., visual similarity, to identify phishing attacks.

\subsection{Visual Similarity Model} \label{sec:visualsimexpt}



\begin{figure*}
\centering
 \includegraphics[width=1.6\columnwidth]{./expt/fig3_VSmodel.pdf}
\caption{Visual similarity model's top-1 matches on the running examples. Top-1 matches are correct for G1, G2, B1 and B2. For D1, the match to a DHL image occurs at 147th rank.} \label{fig:visualphish}
\end{figure*}
\subsubsection{Model Performance}
We train and test the visual similarity model on the extended VisualPhishNet dataset. The details, including hyperparameters and number of epochs, are same as in the original study~\cite{visualphishnet}. Our model has 88.27\% top-1 and 93.61\% top-5 match accuracy. The original study reported 93.25\% top-1 and 96\% top-5 match accuracy. Training the model for more epochs increases its accuracy. To identify whether a webpage is phishing or not, we input the screenshot of a webpage to the model. When the distance score is less than the distance threshold of 8 used in the original study~\cite{visualphishnet}, we consider the input as a phishing page. 
\subsubsection{Running Examples}
The model correctly identifies all five examples as phishing. However, 
from Figure~\ref{fig:visualphish}, we see that the model outputs correct brand for four examples: G1, G2, B1 and B2. The examples G1 and G2 are phishing the Google login page in German language. The model correctly identifies that the webpages are phishing Google. It matches the Google login page in German with the Google login page in English, which shows that the similarity model is robust against variations in the text language. The model learned the visual elements of the Google login page such as placement of the logo, user input box and login button, and detected similar features although the text was in German. Recall that the URL model did not identify G1 as phishing. This is an example where cloud-based phishing impacts the URL model, but not the visual similarity model. 

The model correctly identifies B1 and B2 as  British Telecom (BT) phishing pages. Looking at the top-1 match for B1, we see that the model is robust not only against variations in the text, but also to variations in the visual elements. The size of the person's image  above the `Sign In' text and  the page footer bar are different between the phishing page and the matched page, but the model is correctly able to identify phishing. For B2, the model is able to find its exact match from the dataset.


The model correctly identifies D1 as a phishing page, but the top match for the brand is Facebook (distance 3.1) instead of DHL. In Figure~\ref{fig:visualphish}, we are showing the top-2 match (distance 4.0, also Facebook) because top-1 match image has some aspects of nudity, which is common for many phishing pages in the real world. The model matches D1 to a DHL image at rank 147 with distance 6.92, which is still below the threshold and hence correctly classified as phishing. The model performs poorly on D1 because the D1 page is visually very different from the DHL pages in our training dataset. The entire D1 page is a color image with a relatively large DHL logo in the center and a sky background near top left. In contrast, the DHL training pages have a yellow colored header bar with a smaller DHL logo on the left and smaller font texts throughout the page. This analysis also explains why our top-1 and top-2 matches for D1 are colorful pages with large text logos.

\begin{figure}
\centering
 \includegraphics[width=.7\columnwidth]{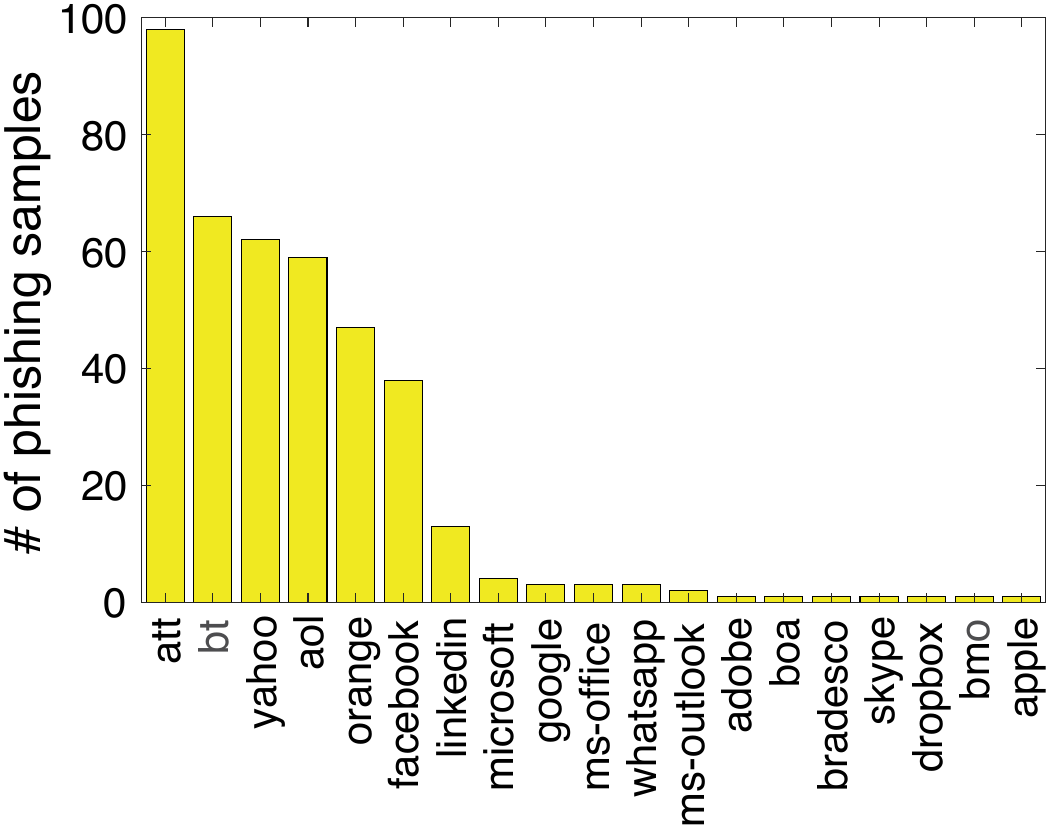}
\caption{Histogram of 405 cloud-based phishing data samples used to test the visual similarity model. \emph{boa} is Bank of America. \emph{bmo} is Bank of Montreal.} \label{fig:largephishdataset}
\end{figure}
\begin{figure*}
\centering
 \includegraphics[width=1.8\columnwidth]{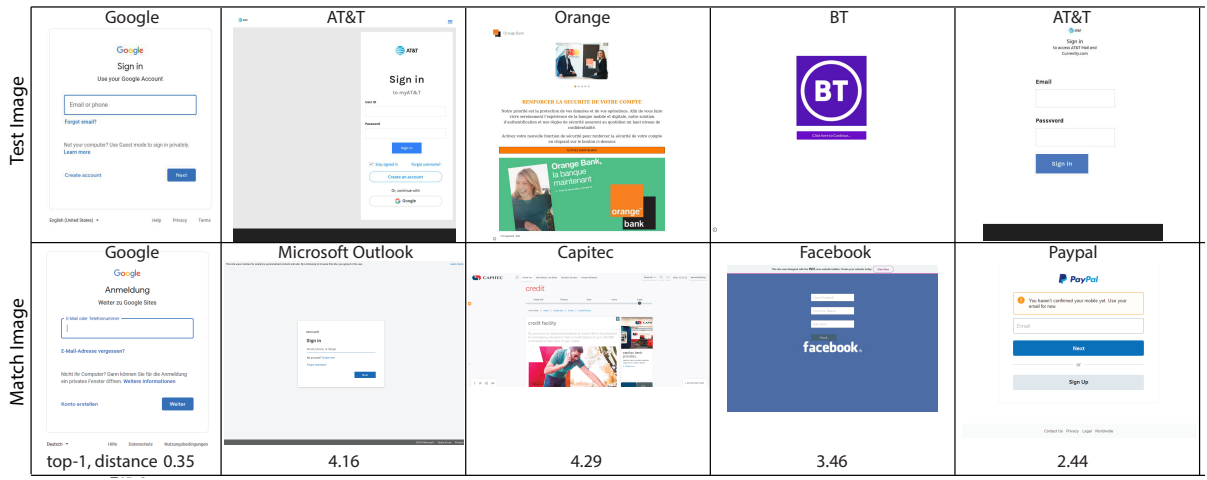}
\caption{Visual similarity model's result on five samples from the cloud-based phishing  dataset. The brand mismatch for 4 out of 5 samples shows that the visual similarity model, when trained on traditional phishing websites and used in isolation, may not be sufficient to detect  cloud-based phishing.} \label{fig:visualphish_largeclouddataset}
\end{figure*}

For additional testing on cloud-based phishing sites, we took 405 page screenshots from the dataset of 455 cloud-based phishing URLs collected during March-June 2022. For the remaining 455$-$405 = 50 images, either the target brand was not part of the VisualPhishNet training set or the image looked very similar to another image in the  dataset, which happens when different URLs point to almost identical looking phishing pages. We selected 405 unique images targeting 19 brands shown in  Figure~\ref{fig:largephishdataset}. 
On these images, our visual similarity model could detect the correct target brand in top-10 matches with 59\% accuracy. The main reason for this relatively poor accuracy is that the test image layout (background and foreground colors, button position and color, size etc.) for many brands is significantly different from the training image layout of the respective brands. This is evident from a relatively high value of the top-1 match distance as shown in  Figure~\ref{fig:visualphish_largeclouddataset}. The figure shows that, for three out of five samples, the top-1 match distance is greater than 3.0 and those matches are incorrect. In the 5th column, the AT\&T image  has been matched to a Paypal image and the distance of 2.44 is  smaller than 3.0  because the two pages are visually similar. This suggests that including cloud-based phishing images in the training of the visual similarity model will improve the model performance on cloud-based phishing. Another point is that, for the same 405 URLs whose screenshots are tested here, the New-3 model discussed earlier detected 69\% of URLs as phishing, which is higher than 59\% of VisualPhishNet.
This supports our original hypothesis: a combination of deep learning models is needed to detect the cloud-based phishing attacks.
\subsection{Logo Model} \label{sec:logodetectexpt}
\begin{figure*}
\centering
\includegraphics[width=1.9\columnwidth]{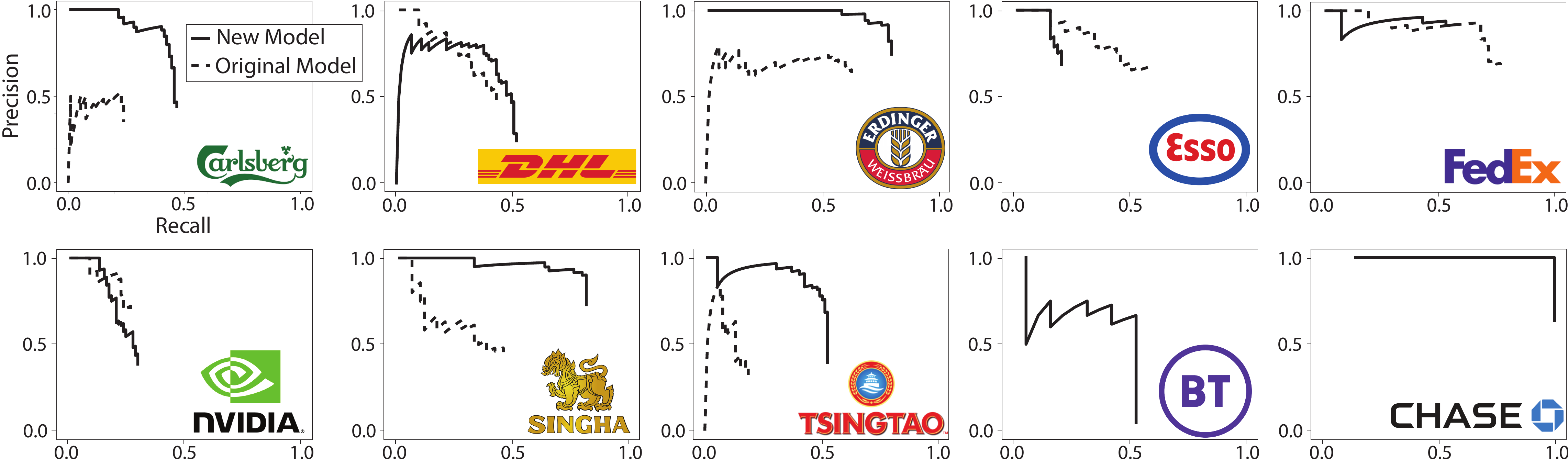}
\caption{Precision vs recall curves from the logo models for 10 selected logos. Dotted line = Original Model trained on the Flickr47 logos, solid line = New Model trained on Flickr47 and extended to include Chase and BT logos} \label{fig:logomodel_prec_recall}
\end{figure*}

\begin{figure}
 \centering
\includegraphics[width=.95\columnwidth]{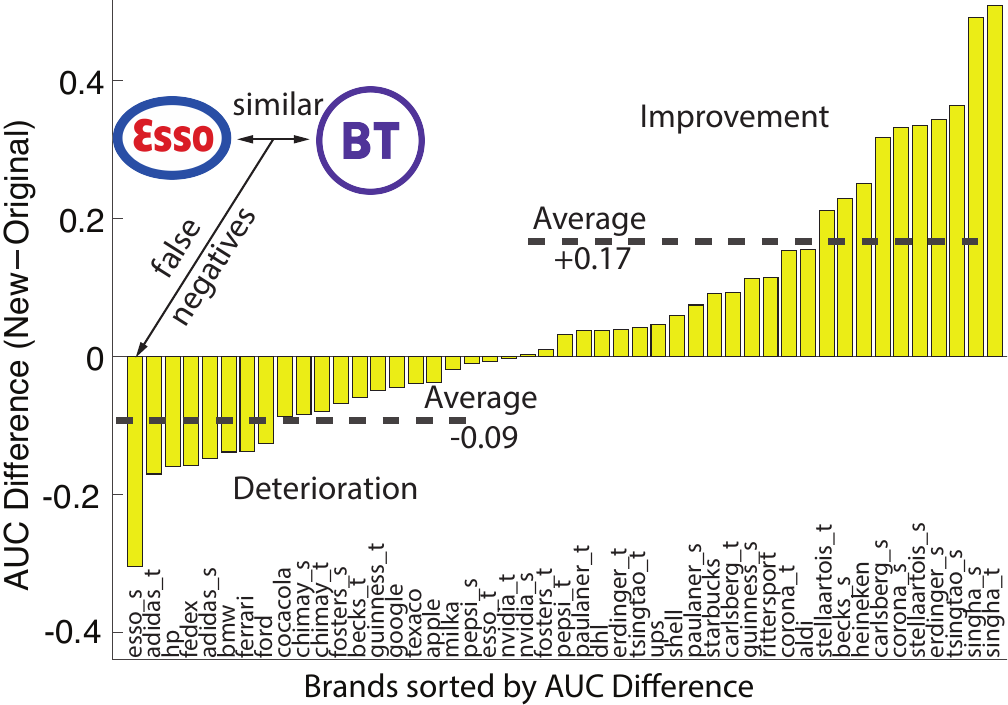}
 \caption{Difference in AUC values between the two logo models. Performance of the new logo model has improved compared to the original model for more than half (26/47) of the brands. Similarity between Esso and BT logos leads to false negatives and AUC Difference of -0.3 for Esso.} \label{fig:aucchange}
\end{figure}

\begin{figure*}
\centering
 \includegraphics[width=1.7\columnwidth]{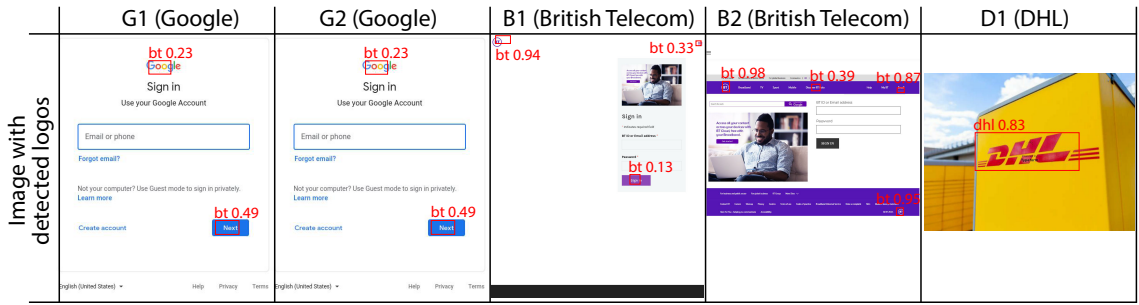}
\caption{Output of our logo model (bounding box, brand name and detection probability) for the running examples.} \label{fig:logo}
\end{figure*}

\subsubsection{Model Performance}
In Section~\ref{sec:design}, we described using transfer learning on a pretrained YOLO2 logo model to include two new brands. During training, we use 0.1 validation split. To evaluate the model performance, we use precision and recall metrics which are commonly used for binary classification models. Given an image, for each logo detected above a detection threshold probability 0.25, the model outputs the following: a bounding box around the logo, name of the most likely brand for the logo, and the probability of that brand. 
If the detected brand label around a region in the image matches the ground truth brand label and the Intersection Over Union (IOU) of the two bounding boxes is at least 0.5, we count it as a true positive for that brand~\cite{rafael2021}, else it is a false positive. If there is no detected brand around a region where ground truth label exists, it is considered as false negative. 
Precision and recall values are calculated based on the number of true positives, false positives, and false negatives. 
We also compute the Area Under the precision vs recall Curve (AUC) for all the brands in the training dataset.
We selected 10 representative brands (our of the total 49 brands) for a detailed view  of precision vs recall curves from the two logo models (Figure~\ref{fig:logomodel_prec_recall}). 

To understand how transfer learning impacted the performance of the new logo model, we plot the AUC difference between the new and original models. Figure~\ref{fig:aucchange} shows that the AUC increased for 27 brands and decreased for 19 brands. The average increase in AUC is more than the average decrease in AUC, which implies that the new model performs better than the original model. 

The decrease in AUC can be due to model overfitting during  transfer learning, where the model is trained for 3000 epochs beyond the original 10000 epochs. The model can overfit to one of the existing brands or to one of the new brands of transfer learning. The former leads to false detection of additional logos of an existing brand on another brand's page (false positive). The latter leads to false detection of the new brand on an existing brand's logo (false negative). Although the detection probability of these additional detections is small ($<$0.3), it leads to an increase in the count of false positives and false negatives and thereby a decrease in precision and/or recall and AUC. 
When we added BT, the similarity between Esso and BT logos led to an increase in false negatives for Esso (detection of BT on Esso images), which led to the maximum AUC drop of -0.3 for Esso. 

\subsubsection{Running Examples}
Note from Figure~\ref{fig:logo} that the logo model correctly identifies the brands for B1, B2 and D1, but not for G1 and G2. The model correctly identifies the DHL logo in D1 with a high probability of 0.83. Here, the model performs well because the logo occupies a relatively large fraction of the image. For B1 and B2, it correctly identifies the BT logo with high probabilities 0.94 and 0.98 respectively. It does so even though the logos occupy a relatively small area of the image. This is because the training images for BT also contained logos that occupied a small fraction of the image. The model does not identify the Google brand in G1 and G2. Instead it misclassifies the brand as BT, which is another text-based logo like Google, although with a low probability of 0.23. Training images for Google contained high resolution logos occupying a relatively large fraction of the images, but G1 and G2 contain small low resolution logos. Hence, the model does not identify the brand correctly. To detect small logos better, we can use YOLOv3~\cite{redmon2018yolov3} or Faster R-CNN~\cite{ren2015fasterRCNN} models. 

To further test the logo model on cloud-based phishing attacks, we chose 69 images out of the 405 cloud-based phishing page screenshots used to test VisualPhishNet in Section~\ref{sec:visualsimexpt}. We chose only 69  because they target Google and BT brands which are included in the training dataset of our logo model. Other 336 images target brands that are not part of the training dataset of our logo model.
Consistent with our previous observation, the Google logos could not be detected due to poor resolution and small size, whereas almost all the images with BT logos are detected correctly albeit with varying probability due to the varying sizes of the BT logos in those images.

\subsection{Combined Model}

Table~\ref{tab:combined} summarizes the outputs of the individual models for the running examples. We note three ways in which combining outputs can improve detection of cloud-based phishing attacks.

First, neither the URL model nor the visual similarity model identifies phishing for all the running examples. The URL model identifies two---B1 and D1---and the similarity model identifies four---G1, G2, B1 and B2---correctly. When we consider both the models together, all five examples are correctly identified as phishing. Each model has its advantages. In our evaluation, inferences time for the URL model is about four times faster, but the accuracy of the visual similarity model is less impacted by cloud-based attacks. In situations where it is important to identify false negatives, i.e., phishing websites that may not get caught  as phishing, running both URL and visual similarity models can add value.

Second, the output of the logo model can enrich the results of the other two models. We can combine the logo model with the URL model to identify the brand being phished. In D1, the URL model identifies phishing with high 0.99 score, and the logo model identifies DHL logo with high 0.83 probability. Combined they identify a phishing attack on DHL. We can use the logo model to identify cases where the visual similarity model does not correctly identify the brand being phished. For D1, the visual similarity model identifies Facebook as the top brand being phished. But because the logo model identifies DHL, we can check the distance score output by the visual similarity model for DHL and whether the score is less than the threshold for phishing. The model outputs a distance of 6.9 
for the closest matched DHL image from the training corpus, and this distance  is less than the threshold of 8 given in the original VisualPhishNet paper. 
Hence, the model considers D1 as a phishing page of DHL, which is correct. 
However, since this match occurs at the 147th rank, the likelihood of this phishing site targeting DHL is considered low.
In general, for the brand identified by the logo model, the similarity model can provide the corresponding distance score. Previous work~\cite{visualphishnet} hypothesized that including logo features could reduce the false positives of a visual similarity model, and our results support that hypothesis.

Third, when we combine outputs from the three models, we can assign a \emph{confidence level} to the result. For example, we can have high confidence that B1 is a phishing attack on BT because the URL model assigns it a high phishing score, visual similarity model outputs a top-1 phishing match for BT, the logo model identifies BT logo with high probability, and the brand identified by logo and visual similarity models match. Assigning a confidence level can help in prioritizing cases that require manual analysis. Let us consider another example G1. Here, the visual similarity model outputs a top-1 phishing match for Google. However, the URL model does not identify phishing and logo model does not detect the Google logo. Hence, we can assign low confidence to the result. Similarly, we can assign low confidence to G2's result. For B2, URL model does not identify phishing but visual similarity model identifies top-1 phishing match for brand BT, and the logo model also identifies BT logo with high probability. Therefore, we can assign it medium confidence. For D1, we assign medium confidence because the URL model identifies phishing with high score and the logo model identifies the DHL logo with high probability, but the visual similarity model identifies phishing on DHL at a lower rank.

\begin{table}[t]
\begin{tabular}{@{}lllll@{}}
\toprule
\multicolumn{1}{c}{\multirow{2}{*}{ID}} & \multicolumn{3}{c}{Individual Model Output} & \multicolumn{1}{c}{\multirow{2}{*}{\begin{tabular}[c]{@{}c@{}}Phishing Detected with\\ Confidence Level\end{tabular}}} \\ \cmidrule(lr){2-4}
\multicolumn{1}{c}{}                    & URL            & Logo        & Similarity   & \multicolumn{1}{c}{}                                                                                                   \\ \midrule
G1                                      & 0.001(No)       & 0.23(No)    & top-1(Yes)   & Yes, low confidence                                                                                                    \\
G2                                      & 0.043(No)    & 0.23(No)    & top-1(Yes)   & Yes, low confidence                                                                                                 \\
B1                                      & 0.997(Yes)      & 0.94(Yes)   & top-1(Yes)   & Yes, high confidence                                                                                                   \\
B2                                      & 0.315(No)       & 0.98(Yes)   & top-1(Yes)   & Yes, medium confidence                                                                                                 \\
D1                                      & 0.999(Yes)      & 0.83(Yes)   & top-1(No)    & Yes, medium confidence                                                                                                 \\ \bottomrule
\end{tabular}
\caption{Combining outputs from the three models: URL, visual similarity and logo detection. \emph{Yes} and \emph{No} are model outputs for \emph{Is phishing detected?} }
\label{tab:combined}
\end{table}

\section{Limitations} \label{sec:limitations}
Compared to prior work, we focus on cloud-based phishing attacks. Through a qualitative analysis, we show how combining deep learning models can help in identifying cloud-based phishing attacks. We validated the performance of the individual models on published datasets, and then examined their performance on cloud-based phishing attacks in the wild. Quantitatively examining performance using larger cloud-based phishing datasets can provide further insights. We studied LSTM, triplet network and YOLOv2 deep learning models. Using advanced versions of the models could improve performance, e.g., YOLOv3 or Faster R-CNN models to detect small logos. In addition to deep learning models for URL, visual similarity and logo analysis, models for analyzing code (HTML/DOM/CSS) could help in detecting cloud-based phishing attacks. Our results show the value of combining model outputs. To generate a combined phishing score, we could use ensemble methods such as averaging and stacking of results from the individual models. 

\section{Discussion and Conclusions} \label{sec:conclusion}
Phishing attacks that use popular cloud web hosting services and apps are rising. Traditional phishing detection methods such as IP reputation monitoring are not effective against these attacks. We investigated the use of deep learning models to identify these attacks. Specifically, we examined LSTM for URL analysis, triplet network for visual similarity analysis and YOLO2 for logo detection. A model that used hand-picked features such as domain similarity index would not adapt to cloud-based phishing attacks. However, deep learning models can automatically learn the best features and can adapt to new class of attacks. The deep learning URL model uses the presence of random character sequences in URLs to identify phishing. However, cloud-based URLs are relatively less random than traditional phishing URLs. Therefore the model is less accurate on cloud-based phishing attacks. In contrast, visual similarity models perform better because they analyze the webpage content as an image. 

While we focused our discussion on cloud-based phishing attacks and their unique challenges, our models are trained on generic datasets that can be used on cloud-based as well as non-cloud based phishing attacks. 
Our evaluation results show that while any one particular model is not sufficient to definitively identify a cloud-based phishing attack, combination of a URL model, a visual similarity model and a logo detection model has the potential to cover a large space of cloud-based phishing attacks. 
By combining model outputs, we can reduce false negatives. We can also assign a confidence level to the results. 
Our results of the LSTM model on 455 cloud-based phishing URLs, visual similarity model on 405 screenshots, and YOLO2 model on 69 screenshots confirm this. 



\begin{figure}
 \centering
\includegraphics[width=.9\columnwidth]{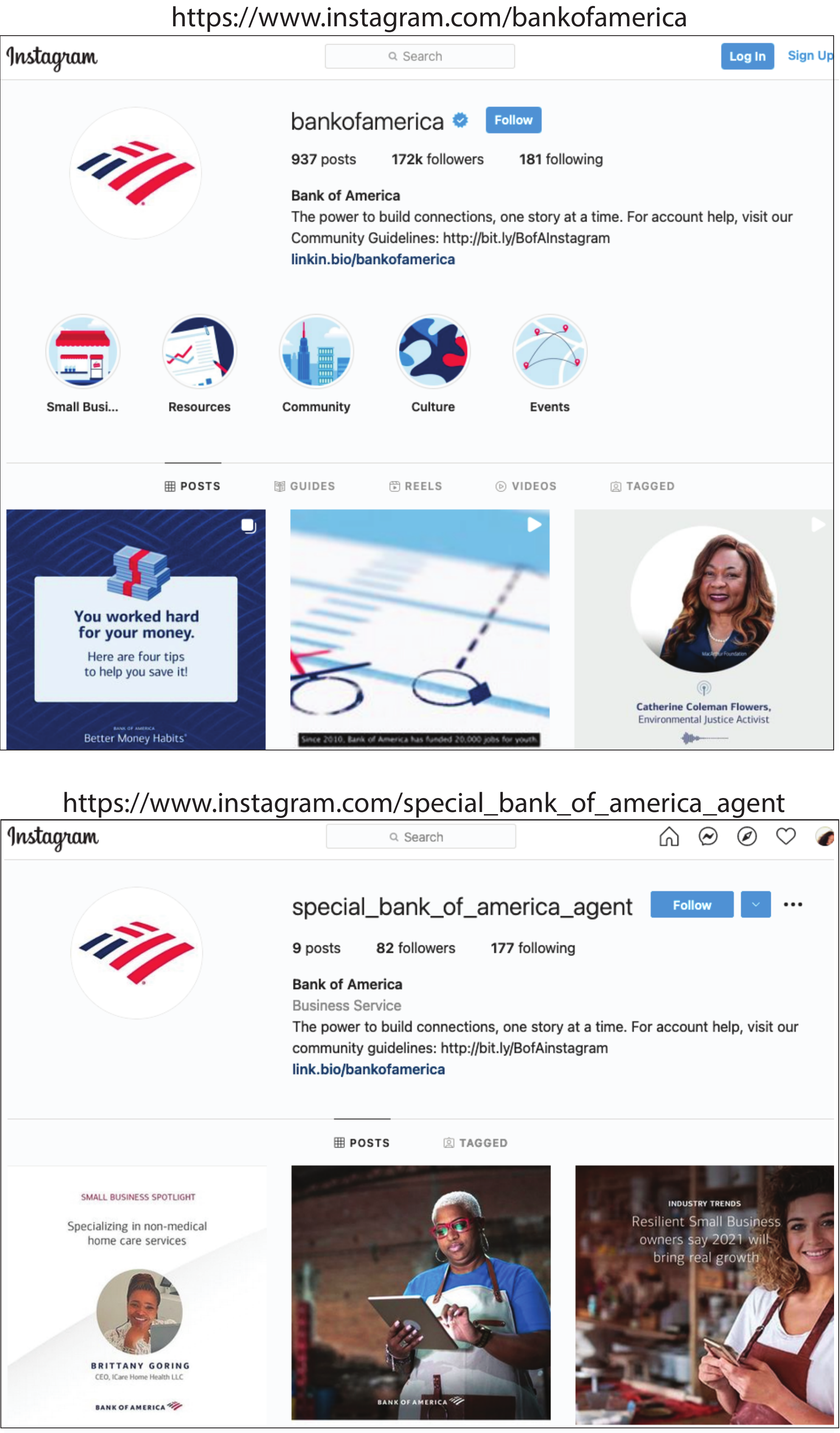}
 \caption{Bank of America's legitimate account (top) and one of the fake accounts (bottom) on Instagram. The URL domain name is the same, the account logos are identical and the bio section of the pages are visually similar. The fake account was taken down recently.} \label{fig:boa_instagram}
\end{figure}
Cloud-based phishing attacks are similar to impersonation attacks on social media. For example, an attacker can create an account instagram.com/\textit{special$\_$bank$\_$of$\_$america$\_$agent} to impersonate a legitimate account instagram.com/\textit{bankofamerica}; see Figure~\ref{fig:boa_instagram}. The attacker then copies the brand logo, a few recent posts (images) and bio from the legitimate account. Similar to cloud-based attacks, IP address monitoring and domain name similarity do not work because the  domain  name for all such accounts is the  social media platform's name. Combining deep learning models for URLs, visual similarity and logo could help in identifying impersonation attacks. However, there are differences between social media phishing and website phishing that need to be addressed. First, the URLs differ only in the account handle part. Hence, there is less information for a URL model to learn. Second, from the visual similarity perspective, the account pages have rigid layouts and are mostly similar across all accounts. Hence, the threshold distance to classify a page as either phishing or  legitimate needs to be different in the social media context. The content of account page i.e. posts can also change daily compared to websites, which may necessitate  retraining of the models. 
Based on these insights, we believe our combined model approach can be extended  to impersonation attacks on social media.


\bibliographystyle{./IEEEtranS}
\bibliography{./references}

\listoffixmes

\end{document}